\documentclass[aps,twocolumn,superscriptaddress,showpacs]{revtex4}
\usepackage{amsmath}
\usepackage{amssymb}
\usepackage{amsfonts}
\usepackage{graphicx}
\usepackage{bm}
\usepackage{epstopdf}

\usepackage{dcolumn}
\usepackage{bm}
\usepackage{graphicx}
\usepackage{color}
\DeclareGraphicsExtensions{. jpg,. pdf, . mps, . png, . eps, . ps, . EPS}

\begin{document}
\def\be{\begin{equation}}
\def\ee{\end{equation}}
\def\bc{\begin{center}} 
\def\ec{\end{center}}
\def\bea{\begin{eqnarray}}
\def\eea{\end{eqnarray}}
\newcommand{\avg}[1]{\langle{#1}\rangle}
\newcommand{\Avg}[1]{\left\langle{#1}\right\rangle}

\title{The  Shannon and the Von Neumann entropy \\of random networks with heterogeneous expected degree}

\author{Kartik Anand}
\affiliation{Technische Universit\"at Berlin, 10623 Berlin, Germany}
\author{ Ginestra Bianconi}  
\affiliation{Department of Physics, Northeastern University, Boston 02115 MA, USA}
\author{Simone  Severini  }
\affiliation{Department of Physics \& Astronomy,University College London, WC1E 6BT London, UK}

\begin{abstract}
Entropic measures of complexity are able to quantify the information encoded in complex network structures.
Several entropic measures have been proposed in this respect. Here we study the relation between the Shannon entropy and the Von Neumann entropy of networks with a given expected degree sequence. We find  in different examples of network topologies  that when the degree distribution contains some heterogeneity, an intriguing correlation emerges between the two entropies. This result seems to suggest that this kind of heterogeneity is implying an equivalence between a quantum and a classical description of networks, which respectively correspond to the Von Neumann and the Shannon entropy.
\end{abstract}
\maketitle
\section{Introduction}

Measures of complexity are central to the investigation of complex networks \cite{Albert2002,Dorogovtsev2003,Boccaletti2006,Dorogovtsev2010} in social, technological and biological contexts. These measures allow us to capture differences and similarities between networks appearing in vastly different contexts, which furthers our understanding of the information encoded in complex networks. In particular, following \cite{Albert2002,Dorogovtsev2003,Boccaletti2006,Dorogovtsev2010} one may define the complexity of a network as a summary of the underlying graph structure. Consequently, the presence of hierarchical organization \cite{Fortunato2010} or a scaling behavior of the degree distribution \cite{Albert2002} are examples of structural features that contribute to the complexity.

Recently, following  information theoretical and statistical mechanics paradigms, several {\it entropic} measures for complexity \cite{Sole2003,Sole2004,BGS,BGS1,Bianconi2008a,Bianconi2008b,Passerini2008,Latora2008,Bianconi2009,Anand2009,Annibale2009, Rovelli2010, Severini2010c,Anand2010,Burda2010,Munoz2010} have been proposed for network structure. These measures have been shown to be extremely successful in quantifying the level of organization encoded in structural features of networks \cite{Sole2004,Gfeller,Features,Severini2010a,Sole2010}. 

Nevertheless, the relationship between the different entropic measures  remains an open question. In a previous paper \cite{Anand2009}, we found insightful connections between the Von Neumann and the Shannon entropy. The Shannon entropy $S$ of a network ensemble is proportional to the logarithm of a typical network in the ensemble. As such, the Shannon entropy of all networks with given structural constraints has the clear interpretation of quantifying the information present in network structures.
The smaller is the entropy of an ensemble, the stronger are the limitations imposed by the constraints and, in some sense, the more optimized is the network. It turns out that the Shannon entropy of an ensemble depends on the {\em number} of constraints that are imposed.
In particular, more constraints imply a smaller entropy.
Additionally, the entropy also qualitatively depends on the {\em type} of constraints. For example, the entropy of scale-free networks    \cite{Bianconi2008a,Bianconi2008b} is much smaller than the entropy of networks with exponential degree distribution also if in both cases only the degree distribution of the network is fixed. This result is able, for example, to quantify how much more information is encoded in such scale-free networks respect to   networks with other degree distributions.

The von Neumann entropy of a network, $S_{VN}$,  was previously introduced  by Braunstein, Ghosh,  and Severini \cite{BGS, BGS1} and later discussed in \cite{Passerini2008} and \cite{Rovelli2010}. This quantity is determined by the spectrum of the graph Laplacian ${\cal L}$ of the network. Specifically, considering networks with $N$ nodes and an average degree of $\avg{k}$, we have that
\begin{equation}
S_{VN}\,=\,-\,{\rm Tr}\,\frac{{\cal L}}{\avg{k}N}\,\log\,\frac{{\cal L}}{\avg{k}N}\,,
\end{equation}
where $\mbox{Tr}\, {\cal L}\,=\,\avg{k}N$.  The authors of \cite{Passerini2008}  demonstrate that the von Neumann entropy may be seen as a measure of regularity in networks, i.e., regular graphs with an equal number of neighbors for all nodes tend to display a higher entropy than those with heterogeneous degree distributions and the same number of links.
Since ${\cal L}$ has eigenvalues $\lambda_i$, for $i=1,\ldots,N$ and at least one eigenvalue is equal to zero, the maximum of the Von Neumann entropy is reached when all other eigenvalues are equal and hence $S_{VN} \le \log(N-1)$. This bound is also saturated by random graphs in the asymptotic limit \cite{Severini2010c}.

In the present paper we explore in greater detail the relation between the Shannon and von Neumann entropies, $S$ and $S_{VN}$, respectively for an ensemble of networks with hidden variables that fix the degree sequence. Our results indicate that for the canonical ensemble of exponential random network, in the limit of large $N$, the entropy $S_{VN} \simeq {\cal O}[\log(N)]$. This rule holds irrespective of the scaling of the average degree $\avg{k}$, and the degree distribution which are responsible for the sub-leading corrections.
Moreover, as soon as we introduce heterogeneity in the expected degree of individual nodes, the von Neumann entropy correlates with the Shannon entropy in a non-trival manner. The heterogeneity of expected degrees induces an intriguing relation between the quantum description of the network captured by the Von Neumann entropy and the classical description of the networks. 
It has been already shown that growing complex networks with heterogeneous feature of the nodes show in their evolution some sort of quantum effects \cite{Bose,W, Complex, Fermi}, in the sense that the evolution of these networks is described well by quantum statistics.

Here we observe that the simple heterogeneity in the degrees induces a correlation between essentially classical and quantum measures of complexity of the networks. It may be valuable to interpret this fact with particular attention to the classical-quantum interface. The suggestion arising from our discussion tells in some way that mixed quantum systems, consisting of ensembles composed by many pure states, have a more classical nature when their von Neumann entropy correlates to the Shannon entropy. In our case, this is exactly when there is some "extra amount of disorder" introduced at the degree level. The hypothesis opens a scenario to further investigate this kind of correlation, with the purpose to study properties of the transition from quantum to classical physics, at least when considering large ensembles of two-level states.

The paper is structured as follows. In Sec. II, we define and evaluate the Shannon entropy for networks ensemble with expected average degree sequence. In Sec. III, we define the Von Neumann entropy of networks in terms of the Laplacian spectra. In Sec. IV, we we show  how the Laplacian spectra can be calulated by the cavity methods. In Sec. V, we compute the Lapalcian spectra of dense networks. In Sec. VI, we use the Effective Medium Approximation for the spectra or sparse networks. Finally, in Sec. VII we show that the Shannon entropy and the Von Neumann entropy of networks with heterogeneous degree distribution correlate in the cases of dense and sparse network ensembles. Conclusions are drawn in Sec. VIII.

\section{Shannon entropy of canonical ensemble with heterogeneous degree distribution}

In general, the entropy is defined as the logarithm of the number of networks in the ensemble. One distinguishes between two types of ensembles; (i) the micro-canonical ensemble, where structural features -- constraints -- are satisfied by all networks in the ensemble. The $G(N,M)$ model \cite{Renyi1959,Janson2000} from the theory of random graphs is a simple example of a micro-canonical ensemble of networks with $N$ nodes, where we dictate that the total number of links is  $M$. The entropy for the micro-canonical ensemble may also be computed via combinatorial methods. (ii) One also defines canonical ensembles, where constraints are satisfied {\it on average}. The complement of the $G(N,M)$ ensemble is the $G(N,p)$ model, which corresponds to the canonical ensemble of networks with $N$ nodes and a fixed probability $p$ for a link to be present. The total number of links $M$ in the $G(N,p)$ ensemble is Poisson distributed with average  $\avg{M}\,=\,p\,N$. 

The parallel between microcanonical and canonical network ensembles can be extended to include more general community, structural and even spatial constraints \cite{Bianconi2008a,Bianconi2008b,Bianconi2009,Anand2009,Annibale2009,Anand2010,Munoz2010}.

In this paper, we focus on the canonical ensemble with fixed expected degree sequence, which is the conjugated canonical ensemble of the configuration model where the degree sequence is fixed. This ensembles is also known as the {\em hidden variable ensembles} and thse have been studied both by statisticians and physicists \cite{Snijders, hv,Boguna,Newman,Bianconi2008a,Bianconi2009,Anand2009}

A network belonging to the canonical ensemble of {\it uncorrelated network} with fixed expected degree sequence may be constructed as follows:
\begin{itemize}
\item For each node $i$, draw its expected degree $q_i$ from the probability distribution $p_{q}$, where $q_i<\sqrt{\avg{q}N}$ and $\avg{q} = \sum_{q^\prime} p_{q^\prime}\,q^\prime$. The upper bound condition for the $q_i$ ensures that they remain uncorrelated across nodes.

\item  Between nodes $i$ and $j$ we add a link with probability
\begin{equation}
p_{ij}=\frac{q_i\,q_j}{\avg{q}\,N}\,.
\label{pij}
\end{equation}
\end{itemize}
In the large network limit $N\to \infty$ the degree $k_i$ of each node $i$  in this ensemble is given by a Poisson variable with mean and variance equal to $q_i$.  Therefore also  the total number of links is a Poisson variable with average  $\avg{k}\to \avg{q}$.
It follows that the average degree of each network realization is a self-averaging quantity for networks in which $<q>=o(N)$, i.e. the expected degree converges in probability to a delta distribution peaked at the expected average degree ($\avg{k}=\avg{q}$).

The  entropy of canonical network ensembles is readily obtained from an appropriate definition of the ensemble \cite{Snijders, hv,Boguna,Newman,Anand2009}.
Each network is defined by its' adjacency matrix ${\cal A}\in\{0,1\}^{N\times N}$, where $a_{ij}=a_{ji}$ describes the presence ($a_{ij}=1$) or absence ($a_{ij}=0$) of a link between nodes $i$ and $j$.
Each network of  the canonical ensemble is assigned a given probability distribution ${\cal P}(\{a\})$ defined over its adjacency matrix $\{a\}$. This is 
\begin{equation}
{\cal P}(\{a\})=\prod_{i<j} p_{ij}^{a_{ij}}(1-p_{ij})^{1-a_{ij}}\,,
\label{eq:PA}
\end{equation}
describing the fact that a link between nodes $i$ and $j$ is present ($a_{ij}=1$) with probability $p_{ij}$  and is absent ($a_{ij}=0$) with probability $(1-p_{ij})$. The {\em Shannon entropy} of such an ensemble
is defined as
\begin{eqnarray}
S&=&-\frac{1}{N}\sum_{\{a\}}{\cal P}(\{a\})\ln {\cal P}(\{a\})\nonumber\\
&=&-\frac{1}{N}\left\{\sum_{i<j}[p_{ij}\log p_{ij}+(1-p_{ij})\log(1-p_{ij})]\right\}.
\end{eqnarray}
Substituting the form $(\ref{pij})$ of the link probability $p_{ij}$ in the canonical ensemble with expected average degrees, we get
\begin{eqnarray}
S&=&\frac{\avg{q}}{2}\log{\avg{q}N}-\frac{1}{N}\sum_{i}q_{i}\ln q_{i}+\nonumber \\
&&-\frac{1}{N}\sum_{i<j}\left(1-\frac{q_{i}q_j}{\avg{q}N}\right)\log\left(1-\frac{q_{i}q_j}{\avg{q}N}\right).
\end{eqnarray}
For convenience, we introduce here the quantity  
\begin{equation}
S_{p}=\int dq\, p_q\, q\,\ln(q)\,.
\label{eq:S_pq}
\end{equation}
This gives us that
\begin{eqnarray}
S&=&\avg{q}\log{\avg{q}N}-S_{p}+\nonumber \\
&&-\frac{1}{N}\int dq \, dq' \,p_q\,p_{q'}\,\left(1-\frac{qq'}{\avg{q}N}\right)\log\left(1-\frac{qq'}{\avg{q}N}\right).\nonumber 
\end{eqnarray}
If we expand the last term in the right hand side, we get the approximate relation
\begin{eqnarray}
S&\simeq &\avg{q}\log{\avg{q}N}+S_{p}-\avg{q}\nonumber \\
&&-\frac{1}{N^2}\int dq \,dq'\, p_q\, p_{q'}\,\left(\frac{qq'}{\avg{q}}\right)^2.
\label{SA}
\end{eqnarray}

\section{Von Neumann entropy}

In quantum mechanics, a \emph{mixed state} is a statistical mixture of \emph{%
pure states}. These are represented by rays in Hilbert space and correspond
to the maximum knowledge which can be acquired about the system.
Mathematically, each quantum state is described by a \emph{density matrix}, 
\emph{i.e. }a positive semidefinite, trace-one, symmetric matrix. The
density matrix of a system with Hilbert space $\mathcal{H}_{N}\cong \mathbb{C%
}^{N}$ is \begin{equation}
\rho =\sum_{i=1}^{N}\omega _{i}|\psi _{i}\rangle \langle \psi _{i}|,
\end{equation} where $%
\omega _{i}$ is a real weight from a probability distribution and $|\psi _{i}\rangle \langle \psi _{i}|$ is the projector
corresponding to the pure state $|\psi _{i}\rangle $. In words, each density
matrix represents a convex combination of pure states, which are then the
extremal points of the set. 

The amount of \textquotedblleft mixedness\textquotedblright\ of a quantum
state is given by the von Neumann entropy, which, in this sense, can be
seen as the quantum analogue of the Shannon entropy \cite{Ohya1993}. 
In information-theoretic terms, the von Neumann entropy quantifies the
incompressible information content of a quantum source, where the signal is
given by pure states; the entropy of a pure state is zero. Additionally, it
has an important operational meaning, being the unique measure of bipartite
entanglement in pure states, via the notion of a reduced density operator.
The connection between quantum mechanics and thermodynamics is expressed by
the fact that $S\left( \rho \right) $ is invariant under Schr\"{o}dinger
evolution, for systems completely isolated from the environment. Indeed,
entropy increases with measurement processes, but \begin{equation}
S\left( \rho \right)
=S\left( U\rho U^{\dagger }\right), \end{equation} for any unitary operator $U$.

The definition of $S$ suggests a direct way to associate a quantum state to
a network, by making use of the graph Laplacian. Importantly, the invariance
under unitary operators guarantees invariance under graph isomorphism. A
discrete analogue of the Laplace-Beltrami operator, the graph Laplacian is a
positive semidefinite symmetric matrix. Once normalized in appropriate way,
it can be treated as a density matrix. Let $G=\left( V,E\right) $ be a simple
undirected graph on $N$ nodes and $L$ links with adjacency matrix $\{a\}$.  From this adjacency matrix we construct the graph {\em Laplacian} ${\cal L}$ where
\begin{equation}
\ell_{ij}\,=\,\delta_{[i;j]}\,k_i\,-\,a_{ij}\,,
\end{equation}
where $k_i\,=\,\sum_{j=1}^N a_{ij}$. The Laplacian operator appears quite frequently in the study of diffusion \cite{Bray1988},  resonance in electric circuits \cite{Fyodorov1999}, and certainly in a myriad of applications in combinatorics and computer science \cite{Mohar1991}. We associate to each node of the network a state in
the standard basis of an Hilbert space of dimension $n$: $1\leftrightarrow
|1\rangle ,...,N\leftrightarrow |N\rangle $. Each link $\{i,j\}$ corresponds
to a pure state $\left( |i\rangle -|j\rangle \right) /\sqrt{2}$. The
associated density matrix is 
\begin{equation}\rho _{\{i,j\}}=\frac{1}{2}\left( |i\rangle
\langle i|+|j\rangle \langle j|-|i\rangle \langle j|-|j\rangle \langle
i|\right).  
\end{equation} The density matrix of the pure state $\left( |i\rangle -|j\rangle \right) /\sqrt{2}$ corresponds to  the Laplacian of a graph with a single link and with $%
N-2$ nodes of degree zero, being $N$ the dimension of the space. The density matrix $\rho _{\{i,j\}}$ is a pure
state. Taking equal weights $\omega _{i,j}$ depending on the number of edges, the matrix 
\begin{equation}
\rho=\frac{1}{\avg{k}N}\sum_{E}\rho _{\{i,j\}}=\frac{1}{\avg{k}N}{\cal L}   \label{lapma}
\end{equation}%
is just the Laplacian of the graph, but adjusted to have unit trace. In this
expression, the weights may be chosen arbitrarily, with the only constraint  $\sum_{\{i,j\}\in E}\omega _{i,j}=1$.

Since the smallest eigenvalue of the Laplacian is zero, we expect that
$S_{VN}\in[0,\log(N-1)]$. The maximum $S_{VN}$ is  attained by the complete graph on $N$ nodes, and in the thermodynamic limit by any Poisson random graph \cite{Severini2010c}. The
minimum is clearly for pure states. 

The Von Neumann entropy on suitable rotation of the Laplacian matrix about its eigenvector basis can be expressed as
\begin{equation}
S_{VN}\,=\,\ln(N)\,-\,\int {\rm d}\lambda\,m(\lambda)\,\frac{\lambda}{\avg{k}}\,\ln\left(\frac{\lambda}{\avg{k}}\right)\,,
\label{sub}
\end{equation}
where the term $m(\lambda)$ denotes the density of eigenvalue state $\lambda$ of the Laplacian.
Therefore the Von Neumann entropy has a leading term of order $\log (N)$ and a subleading terms determined by the spectrum of the Laplacian.
Apart from issues arising
from cospectrality, where nonisomorphic graph can have the same entropy, it
is also worth observing that the expansion properties of the graph do not
seem to be responsible for the behaviour of the entropy, since the fact that the
eigenvalue gap \textquotedblleft compresses\textquotedblright\ the
eigenvalues towards zero does not have other direct implications.
In fact we  observe that the second term on the right hand side of Eq. $(\ref{sub})$  is dominated by the behavior of the density of states $m(\lambda)$ for large values of the eigenvalues $\lambda$'s. Therefore, the large eigenvalues of the graph Laplacian,  that determine the fast temporal scales of the diffusion dynamics on the network, gives the largest contribution to the Von Neumann entropy.

The Von Neumann entropy is defined on every single graph. Nevertheless, it is instructing to evaluate the Von Neumann entropy of ensembles of graphs with different degree distribution. This helps to understand the impact
of the heterogeneity of the network on this new quantity relating networks to quantum states.
In canonical network ensembles the Laplacian spectra is self-averaging as long as $\avg{k}={\cal o}(N)$, meaning that the spectra of a network of this ensemble  will approach the average spectra of the network in the ensemble in the large $N$ limit.
We note also that since in the canonical ensemble the average degree $\avg{k}$ is a self-averaging quantity, the Von Neumann entropy of large networks in the canonical network ensembles can be written also as
\begin{equation}
S_{VN}\,=\,\ln(N)\,-\,\int {\rm d}\lambda\,m(\lambda)\,\frac{\lambda}{\avg{q}}\,\ln\left(\frac{\lambda}{\avg{q}}\right)\,.
\label{sub2}
\end{equation}

\section{Laplacian Spectra of Complex Networks  }
As mentioned in the previous section, the Laplacian spectrum of a network in the canonical network ensembles is self-averaging. Thus, the density of eigenvalues $\lambda$ is given by
\begin{equation}
m(\lambda)\,=\,\lim_{N\,\to\,\infty}\frac{1}{N}\sum_{i=1}^N\overline{\delta(\lambda\,-\,\lambda_i)}^{\cal A}\,,
\end{equation}
where $\{\lambda_i\}$ is the sef of the eigenvalues of ${\cal L}$ and the over-line $\overline{(\ldots)}^{\cal A}$
denotes the average over the probability ${\cal P}(\{a\})$. This is given by Eq. $(\ref{eq:PA})$, for networks in the canonical ensemble. To calculate this object in the large $N\,\to\,\infty$ limit, one introduces \cite{Edwards1976} the partition function
\begin{equation}
\label{eq: Partition function}
Z(\lambda)\,=\,\int \prod_{i=1}^N {\rm d}\phi_i \exp\Big( -\frac{{\rm i}}{2}\lambda_\epsilon\sum_{i=1}^N \phi_i^2\,+\,\frac{{\rm i}}{2}\sum_{i,j=1}^N \phi_i\,\ell_{ij}\,\phi_j\Big)\,,
\end{equation}
where $\lambda_\epsilon\,=\,\lambda\,+\,{\rm i}\,\epsilon$. The introduction of a small imaginary $\epsilon$ is to ensure the integrals are convergent and well defined. In terms of $Z$, we have
\begin{equation}
m(\lambda)\,=\lim_{N\,\to\,\infty}\,-\frac{2}{N\,\pi}\,{\rm Im}\, \frac{\partial}{\partial \lambda}\overline{\log\,Z(\lambda)}^{\cal A}\,.
\label{eq: DOS - partition function}
\end{equation}
To facilitate the averaging over the canonical ensemble, one uses the replica trick and writes 
\begin{equation}
\overline{\log Z(\lambda)}^{\cal A}\,=\,\lim_{n\,\to\,0}\frac{1}{n}\log\overline{Z(\lambda)^n}^{\cal A},
\end{equation}
 where
\begin{widetext}
\begin{equation}
\overline{Z(\lambda)^n}^{\cal A} \,=\,\int \prod_{a=1}^n\prod_{i=1}^N {\rm d}\phi_{ia} \exp\Bigg( -\frac{{\rm i}}{2}\lambda_\epsilon\sum_{a=1}^n\sum_{i=1}^N \phi_{ia}^2\,+\,\frac{1}{2}\sum_{i,j=1}^N \frac{q_i\,q_j}{\avg{q}\,N}\Big[\exp\Big(\frac{{\rm i}}{2}\sum_{a=1}^n(\phi_{ia}\,-\,\phi_{ja})^2\Big)\,-\,1\Big]\Bigg)\,.
\label{eq:avg replicated partition function}
\end{equation}
\end{widetext}
This result is independent of scaling assumptions on $\avg{q}$. In what follows, we consider the cases where $\avg{q}$ scales with $N$, i.e., a dense network, and where $\avg{q}\simeq {\cal O}[1]$, separately.

\section{Dense networks}

For dense network ensembles with divergent average degree, i.e., when $\avg{q}\to \infty $ as $N\to \infty\,$, it is possible to  further simplify Eq. (\ref{eq:avg replicated partition function}), by expanding the inner exponential, and obtain
\begin{widetext}
\begin{equation}
\overline{{\cal Z}(\mu)^n}^{\cal A} \,=\,\int \prod_{a=1}^n\prod_{i=1}^N {\rm d}\phi_{ia} \exp\Bigg( -\frac{{\rm i}}{2}\lambda_\epsilon\sum_{a=1}^n\sum_{i=1}^N \phi_{ia}^2\,+\,\frac{{\rm i}}{4\,N} \sum_{i,j=1}^N\,\sum_{a=1}^n\,x_i\,x_j\,(\phi_{ia}\,-\,\phi_{ja})^2\Big]\Bigg)\,,
\label{eq: avg dense replicated partition function}
\end{equation}
\end{widetext}
where $x_i = q_i/\avg{q}$. We now consider two further sub-cases for the distribution $p_q$ of the $q_i$ expected degrees.

\begin{figure}
\begin{center}
\includegraphics[width=.7\columnwidth]{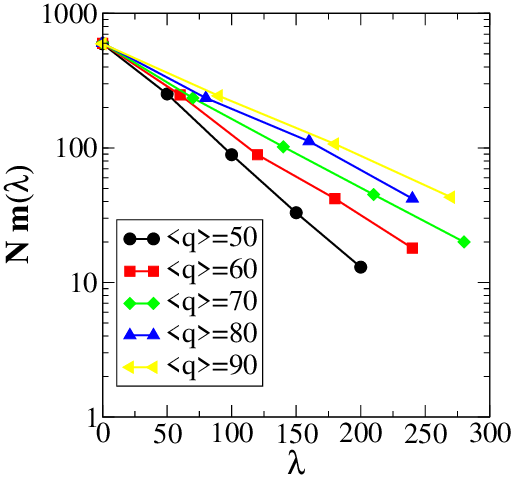}
\end{center}
\caption{(Color online) Laplacian spectra of dense networks with exponential expected degree distribution. The data are obtained from networks of size $N=1000$; the number of links and $p_q$ are distributed exponentially with average $\avg{q}=50,60,70,80,90$. The data are averaged over 20 network realization.}
\label{Dense}
\end{figure}

\begin{itemize}

\item{\it Poisson networks.}
In this case, $q_i\,=\,\avg{q}$ and hence $x_i\,=\,1$. Following \cite{Bray1988}, the Laplacian matrix ${\cal L}/{\avg{q}}$ has one eigenvalue equal to unity and the remaining $N-1$ eigenvalues equal to zero, i.e.,
\begin{equation}
m(\lambda)\,=\,\frac{1}{N}\delta_{[\lambda,1]}\,+\,\frac{N\,-\,1}{N}\delta_{[\lambda,0]}\,.
\label{eq: density of state dense 1}
\end{equation}
Putting this result into Eq. (\ref{sub2}) we obtain that
\begin{equation}
S_{VN}=\ln(N)\,+o(1)\,.
\end{equation}

\item{\it Exponential networks.}
For uncorrelated exponential network ensembles, the probability that a randomly selected node $i$ has expected degree $q_i = q$ is given by
\begin{equation}
p_q=\frac{1}{\avg{q}}e^{-q/\avg{q}}\,.
\end{equation}
Denoting $x_i = q_i/\avg{q}$, the probability that node $i$ has an associated expected degree $x$ is given by $\pi(x)=e^{-x}$. The eigenvalues $\lambda$ are given by $\lambda=x_i$ with degeneracy $N\,\pi(x_i)-1$ if $N\,\pi(x_i)>2$, while the remaining $K$ eigenvalues are equal to zero. For a dense exponential network of size $N$, the value of $K$ is given by $K=\ln N-\ln 2$. Therefore, the density of states is equal to 
\begin{eqnarray}
\nonumber m(\lambda) &=& \left[\pi(\lambda)-\frac{1}{N}\right]\Theta(K-\lambda)(1-\delta_{[\lambda,0]})\\
&+& \frac{K}{N} \delta_{[\lambda,0]}\,.
\label{eq: density of state dense}
\end{eqnarray}
It follows that, in the limit of large $N$, the  spectrum $m(\lambda)$ is formed by a peak for $\lambda=0$ and a bulk spectral density that is decaying  exponentially with a behavior reminiscent of the degree distribution.  
So, the Von Neumann entropy is
\begin{eqnarray}
S_{VN}\,&=&\,\ln(N)\,-\,\int_0^{K} {\rm d}x\, e^{-x}\,\,{x}\,\ln(x)\,+o(1),\nonumber \\
\label{eq:S_VN}
\end{eqnarray}
\end{itemize}
In Fig. $\ref{Dense}$, we represent the spectrum of large dense networks with exponential degree distribution showing the exponentially decaying density of states.
Therefore the Von Neumann entropy of the exponential dense random ensembles has a term  $\log(N)$ and a subleading term equal to the entropy of the degree distribution $S_{p}$ given by Eq. (\ref{eq:S_pq}),  
\begin{eqnarray}
S_{VN}&=& \ln(N)-S_{p}\,.
\end{eqnarray}

\section{Sparse network ensembles}

For sparse network ensembles, where $\avg{q}\sim{\cal O}(1)$, the analysis of the previous section does not hold. In particular, we cannot expand the inner exponential in Eq. (\ref{eq:avg replicated partition function}) in terms of large $\avg{q}$. 
Predicting Laplacian spectra of sparse graphs is a problem that up to date has been only solved numerically, by the use of population dynamics techniques or message passing techniques.
In this paper, in order to determine the behavior of the tail of the Laplacian spectrum we make use of the Effective Medium Approximation \cite{Biroli1999}, which gives a fairly good approximation of the bulk properties of the spectrum.
In Appendix \ref{Ap}, we also show how to derive the Effective Medium Approximation from the exact calculations of the spectrum by population dynamics.

Assuming that the expected degree $q$ can assume any real values drawn from a $p_q$ distribution, the density of states $m_E(\lambda)$ in the Effective Medium Approximation  is given by
\begin{equation}
m_E(\lambda)= \frac{1}{\pi} \int {\rm d}q\, \,p_q\, \mbox{Im}\, \frac{1}{\lambda_\epsilon\,+\,q h_{EMA}(\lambda)},
\label{u}
\end{equation} 
with the variable $h_{EMA}$ satisfying the self consistent equation
\begin{equation}
h_{EMA}=-\int {\rm d}q\, \frac{q\,p_q}{\avg{q}}\frac{1}{\lambda_\epsilon+q h_{EMA}(\lambda)}-1.
\label{u3}
\end{equation}

\subsection{Sparse Poisson random networks}
For the case of the $G(N,p)$ ensemble the spectra is given by 
\begin{equation}
m_E(\lambda)=\frac{1}{\pi}\,\mbox{Im} \,\frac{1}{\lambda_\epsilon+q h_{EMA}},
\label{r}
\end{equation}
with  $q=pN$, and 
\begin{equation}
h_{EMA}=-\frac{1}{\lambda_\epsilon - 1 +qh_{EMA}}-1
\end{equation}
The solution to this equation is
\begin{eqnarray}
qh_{EMA}&=&\frac{1}{2}\left[-(q-1+\lambda_\epsilon)\right.\pm
\nonumber \\
&& \left.i\sqrt{4q\lambda_\epsilon - (1+\lambda_\epsilon-q)^2}\right].
\label{s}
\end{eqnarray}
Plugging this solution into Eq. $(\ref{r})$, we see that in order to have a positive density in Eq. $(\ref{s})$, the effective field $h_{EMA}$ must be complex with negative imaginary part.
It follows that in the limit $\epsilon \to 0$,
\begin{equation}
m_E(\lambda)=\frac{2}{\pi}\frac{\sqrt{4q\lambda-(1-\lambda-q)^2}}{(q-1-\lambda)^2+(q-1+\lambda)^2-4\lambda q}.
\end{equation}
We can see that the spectrum of the Laplacian is not expected to show clear correlation with the degree distribution of Poisson networks.
As a consequence of this, the Von Neumann entropy of Poisson sparse networks is not correlated with the Shannon entropy of the networks or the Shannon entropy of their degree distributions.

\subsection{Sparse Networks with heterogeneous expected degree}

\begin{figure}
\begin{center}
\includegraphics[width=.7\columnwidth]{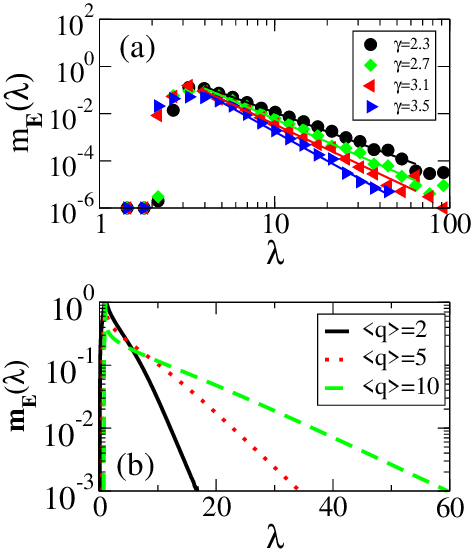}
\end{center}
\caption{(Color online) Laplacian spectra predicted by the Effective Medium Approximation for graphs with power-law  (Panel (a)) and exponential (Panel (b)) distribution of the expected degrees. The predicted spectrum for large eigenvalues $\lambda$  has a power-law behavior for scale-free networks and an exponential behavior for exponential networks.}
\label{Theory}
\end{figure}

\begin{figure}
\begin{center}
\includegraphics[width=.7\columnwidth]{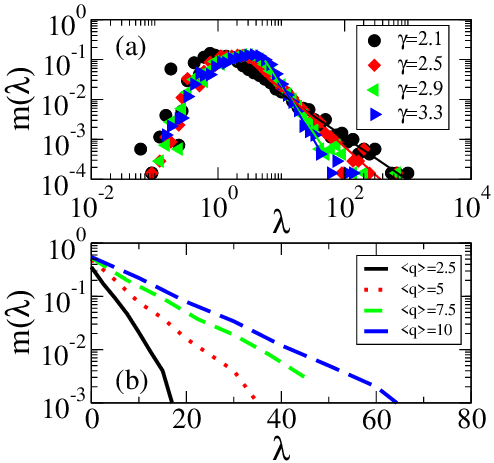}
\end{center}
\caption{(Color online)  Spectra of single networks with power-law (Panel (a)) and exponential (Panel (b)) expected degree distribution. As predicted by the Effective Medium Approximation, the spectra of the Laplacian matrices of power-law networks has a power-law tail, while the spectra of networks with expected exponential degree distribution have exponential tail. The data in Panel (a) are obtained from single networks of size $N=10^4$; the data in Panel (b) are obtained from 20 network realizations of size $N=10^3$.  }
\label{Sparse}
\end{figure}

We have numerically solved Eqs. $(\ref{u})-(\ref{u3})$ defining the spectra of networks with heterogeneous expected degreee in the Effective Medium Approximation.
In Fig. $\ref{Theory}$, we report the predicted spectra for space exponential networks with degree distribution
\begin{equation}
p_E(q)=\frac{1}{\avg{q}}e^{-q/\avg{q}}
\end{equation}
and for power-law networks with expected degree distribution
\begin{equation}
p_{SF}(q)=m^{\gamma-1}(\gamma-1)q^{-\gamma}.
\end{equation}
In the case of large eigenvalues, the spectral density follows a power-law distribution for scale free networks and an exponential distribution for exponential networks. The theoretical expectation are confirmed by direct diagonalization of the Laplacian matrices reported in Fig. $\ref{Sparse}$. Indeed, it is explicitly shown that the tail of the spectrum of the Laplacian has an exponential behavior for networks with exponential degree distribution, and a power-law tail for networks with a power-law expected degree distribution. In Fig. $\ref{Theory}$, we have numerically integrated the Effective Medium equations.

\begin{figure}
\begin{center}
\includegraphics[width=.9\columnwidth]{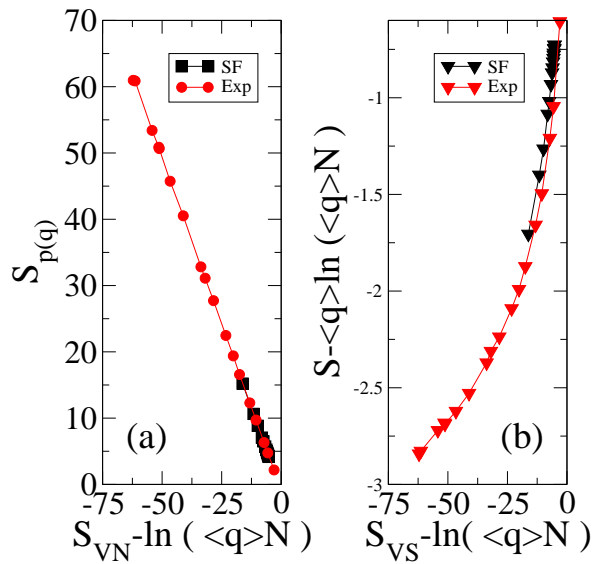}
\end{center}
\caption{(Color online) Comparison between the Von Neumann entropy $S_{VN}$, the Shannon entropy $S$, and the entropy of the expected degree distribution $S_{p}$. These are for networks with power-law and exponential degree distribution, as a function of the average degree and the value of the power-law exponent $\gamma$ of the scale-free expected distribution. For every value of the Shannon entropy, an average is performed over 20 network realizations with exactly the same Shannon entropy.}
\label{Entropies}
\end{figure}

\section{Correlation between the Shannon and the Von Neumann entropy in sparse networks}
The prediction of the Effective Medium Approximation indicates that as soon as the heterogeneity of the degree is introduced the Von Neumann entropy is dominated by the tail of the Laplacian spectrum. Since this tail has in the first approximation the same distribution as the expected degree, the Von Neumann entropy is linearly correlated with the entropy associated with the expected degrees.
Moreover, since the Shannon entropy is also related to the entropy associated to the expected degrees, this relation is reflected in a clear correlation between the Shannon and the Von Neumann entropy.
In Fig. $\ref{Entropies}$ we show that the expected results about the eigenvalues of the Laplacian of scale-free and exponential sparse network are confirmed by the simulation data. The data are obtained by comparing the Shannon and the Von Neumann entropy, and the entropy associated to a given expected (exponential or power-law) degree distribution. All the networks have size $N=1000$ and the different points correspond to network with different average degree for the exponential case, and network with different power-law exponent for power-law networks.

\section{Conclusions}
We have studied the relation between the von Neumann entropy and the Shannon entropy of networks with given expected degree sequence. We report facts occuring for a variety of networks, including dense and sparse networks. We have seen that, when a certain  heterogeneity of the expected degree sequence is introduced, an intriguing connection appears between Von Neumann and Shannon entropy. This connection is missing for networks with homogeneously distributed average degrees.

More than trying to interpret immediately the observed connection, it is worth instead to formulate an open scenery, with the purpose of directing a future line of research. Essentially, it is legitimate to ask whether the Shannon and the von Neumann entropy are correlated for mixed states exhibiting a certain degree of classicality. In this light, it is plausible that the relation between these two entropies can shed a new light into discussions about the quantum-to-classical transition. The specific aspect of this framework does not have the potential to address problems associated to decoherence of pure states, but the classical characteristics of mixtures with many components.
\\
\\
{\it Acknowledgments.} Part of this work has been done while GB was visiting the Department of Physics and Astronomy at University College London. KA acknowledges the support of by the Deutsche Forschungsgemeinschaft through the SFB 469 ``Economic Risks". SS is supported by a Newton International Fellowship.

\appendix

\section{\label{Ap} Derivation of the Effective Medium Approximation from the Replica Method}

In sparse network ensemble, it is possible to find a solution for a matrix partition function, Eq. $(\ref{eq:avg replicated partition function})$, by introducing a functional order-parameter. This approach, first introduced in the context of dilute spin-glass systems  \cite{Monasson1998}, has had wide success in a host of different contexts. Formally, defining $\vec\phi=(\phi_1,\ldots,\phi_n)$, we introduce
\begin{equation}
c_q(\vec\phi) = \frac{1}{N}\sum_{i=1}^{N} \delta_{[q,q_i]} \, \prod_{a=1}^n\Bigg[\delta\Big(\phi_a\,-\,\phi_{ia}\Big)\Bigg]\,.
\label{eq: order parameter}
\end{equation}
Following standard arguments, we obtain the following functional saddle-point expression for the partition function
\begin{equation}
\overline{Z(\lambda)^n}^{\cal A} \,=\,\int \prod_{q>0}{\cal D}c_q(\vec\phi)e^{N\,\Psi(\{c_q\};\,\lambda)}\,,
\label{eq: saddle point}
\end{equation}
where $\Psi(\{c_q\};\,\lambda) = S_0+S_1+S_2$ and 
\begin{widetext}
\begin{eqnarray}
S_0 &=& -\sum_{q>0} p_q \int {\rm d}\vec\phi\,c_q(\vec\phi)\,\log c_q(\vec\phi)\,,\\
S_1 &=&  -\frac{{\rm i}}{2}\lambda_\epsilon \sum_{q>0} p_q \int {\rm d}\vec\phi\,c_q(\vec\phi)\,\vec\phi\cdot\vec\phi\,,\\
S_2 &=& \frac{1}{\avg{q}}\sum_{q,r>0}p_q\,p_r\,q\,r \int {\rm d}\vec\phi\,{\rm d}\vec\psi\,c_q(\vec\phi)\,c_r(\vec\psi)\,\Big[\exp\Big(\frac{{\rm i}}{2\,}\sum_{a=1}^n(\phi_{a}\,-\,\psi_{a})^2\Big)\,-\,1\Big]\,.
\end{eqnarray}
\end{widetext}
As before, $p_q$ is the probability distribution for the expected degrees. From this result together with Eq. (\ref{eq: DOS - partition function}), we note that the density of eigenvalues is given by the variance of the functional $c_q(\vec\phi)$, i.e.,
\begin{equation}
\label{eq:rhocq}
m(\lambda)\,=\,\lim_{n \to 0}\frac{1}{n \pi}\,{\rm Re} \sum_{q>0} p_q \,\langle \vec\phi\cdot \vec\phi \rangle_{c_q}\,,
\end{equation}
where the angled brackets refer to the average over the saddle-point $c_q(\vec\phi)$ measure. This is given by the self-consistent equation
\begin{equation}
c_q(\vec{\phi}) = \exp\left(-\frac{{\rm i}}{2}\lambda_\epsilon \vec\phi\cdot\vec\phi - q \Big[ 1 - \widehat{c}(\vec{\phi}) \Big]\right)\,,
\end{equation}
where
\begin{equation}
\widehat{c}(\vec{\phi}) = \frac{1}{\langle q \rangle}\sum_{r>0} p_r\,r\,\int {\rm d}\vec\psi\,c_r(\vec\psi)\,\exp\left( \frac{{\rm i}}{2\,} \sum_{a=1}^n (\phi_a - \psi_a)^2 \right)\,.
\end{equation}
It is not possible to obtain an analytic closed expression for the $c_q$ measure without making further assumptions. The most general ansatz that one can make here, and expect to hold, is that the measures are invariant under replica symmetry. This amounts to write $c_q$ as a {\it superposition} of Gaussian measures. In particular, we write that
\begin{equation}
c_q(\vec\phi) = \int {\rm d}h_q P(h_q) \prod_{a=1}^n e^{-\frac{i}{2}h_q\phi_a^2}\left(\sqrt{\frac{h_q}{2\pi}}\right)^n\,,
\end{equation}
and
\begin{equation}
\widehat{c}(\vec\phi) = \int {\rm d}\widehat{h} \widehat{P}(\widehat{h})\prod_{a=1}^n e^{-\frac{i}{2}\widehat{h}\phi_a^2}\,.
\end{equation}
Thus, the Gaussian measures are parametrized by the {\it complex} variances $1/h_q$ and $1/\widehat{h}$, which are solved for to obtain the recursive relationship
\begin{widetext}
\begin{equation}
\label{popdy1}P(h_q) = \sum_{r>0} \frac{e^{-q}q^k}{k!}\int {\rm d}\widehat{h}^1 \ldots \int {\rm d}\widehat{h}^k \prod_{\ell=1}^k \left[ \widehat{P}(\widehat{h}^\ell) \right]\delta\left(h_q - \Big[\sum_{\ell=1}^k \widehat{h}^k + \lambda_\epsilon\Big]\right)\,,
\end{equation}\end{widetext}
and
\begin{equation}
\label{popdy2}\widehat{P}(\widehat{h}) = \sum_{q>0}\frac{p_q\,q}{\langle q \rangle}\int {\rm d}h_q P(h_q) \delta \left( \widehat{h} - \frac{h_q}{\,h_q - 1} \right)\,.
\end{equation}
These two equations may be numerically solved, in the general case, using a population dynamics algorithm, as described in the next appendix. To conclude, the density of states is given by
\begin{equation}
m(\lambda)=\frac{1}{\pi}\sum_{q>0} p_q\,\mbox{Im}\,\frac{1}{h_q}.
\label{m}
\end{equation}

\subsection{The population dynamics algorithm}
Eqs. $(\ref{popdy1})$ and $(\ref{popdy2})$ can be solved by a population dynamics algorithm to find the distribution $\widehat{P}(\widehat{h})$. We summarize the algorithm as follows:

{\bf algorithm } PopDyn($\{\widehat{h}_{q_i}\}$ of $M$ fields)
{\bf begin}
{\bf do }
\begin{itemize}
\item choose a field $\widehat{h}_{q_i}$ relative to a node $i$ of the network.
 \item draw $k$ from a Poisson distribution with probability $e^{-q_i}q_i^k/k!$.
\item select $n=1,\ldots,k$ fields  $\hat{h}_{q_j}^n$ relative to nodes $j$, chosen with probability proportional to $q_j$.
\bea
{\hat{h}}_{q_i}:&=&\lambda_{\epsilon}-\sum_{n=1}^k\frac{1}{ \hat{h}^n_{q_j}-1}-k\,.
\label{A14}
\eea
\item update probability distribution function $\widehat{P}(\widehat{h})$.
\end{itemize}
{\bf while} ($\widehat{P}(\widehat{h})$ not converged)\\
{\bf return}\\
{\bf end}\\
\subsection{Derivation of the Effective Medium Approximation }
Finally, this population dynamic algorithm can be approximated by the Effective Medium Approximation (EMA) equations. Here we describe the derivation of the equations presented in the main text from the population dynamics algorithm. 
In the EMA, we average Eq. $(\ref{A14})$ over the probability that node $i$ with hidden variable $q_i$ has $k$ links, and also over the hidden variables $q_j$ of node $i$'s neighbors. 
Therefore using Eq. ($\ref{pij}$) and that the expected degree of node $i$ is $\avg{k_i}=q_i$, we get for the average of Eq. $(\ref{A14})$, the following expression;
\begin{equation}
\hat{h}_q=\lambda_{\epsilon}-q-q\int {\rm d}q^{\prime}\, \frac{q^\prime\,p_{q^\prime}}{\avg{q}}\frac{1}{\hat{h}_{q^{\prime}}-1}.
\label{uno}
\end{equation}
Therefore, if we define
\begin{equation}
{h}_{EMA}=-\int {\rm d}q\,\frac{q\,p_q}{\avg{q}}\frac{1}{\hat{h}_{q}-1}-1,
\label{due}
\end{equation}
using Eq. $(\ref{uno})$, the self-consistent equation for ${h}_{EMA}$ defined by Eq. $(\ref{due})$ is given by
\begin{equation}
{h}_{EMA}=-\int {\rm d}q\,\frac{q\,p_q}{\avg{q}}\frac{1}{\lambda_{\epsilon} - 1 + q h_{EMA}} - 1.
\end{equation}
Finally, inserting in Eq. $(\ref{m})$ the Effective Medium Approximation for the fields $h_q$, i.e., $h_q=\lambda_{\epsilon}+qh_{EMA}$ the density of states $m_E$ in the Effective Medium Approximation is given by 
\begin{equation}
m_E(\lambda)= \frac{1}{\pi}\int {\rm d}q\,\, p_q\, \mbox{Im} \frac{1}{\lambda_{\epsilon} + qh_{EMA}(\lambda)}.
\label{u2}
\end{equation}

\end{document}